\documentclass[aps,tightenlines,showpacs]{revtex4}
\usepackage{graphicx}
\usepackage{float}
\usepackage[T1]{fontenc}
\usepackage[latin1]{inputenc}
\usepackage{amssymb}
\include{epsf}
\epsfverbosetrue

\newcommand{\bea}{\begin{eqnarray}}
\newcommand{\eea}{\end{eqnarray}}

\makeatother
\begin{document}

\title{Energy Flow in Acoustic Black Holes}

\author{K. Choy, T. Kruk, M.E. Carrington and T. Fugleberg}

\affiliation{Department of Physics, Brandon University,\\
Brandon, Manitoba, R7A 6A9 Canada\\
and Winnipeg Institute for Theoretical Physics,\\
Winnipeg, Manitoba, Canada}

\author{J. Zahn, R. Kobes, G. Kunstatter}

\affiliation{Department of Physics, University of Winnipeg,\\
Winnipeg, Manitoba, R7A 6A9 Canada\\
and Winnipeg Institute for Theoretical Physics,\\
Winnipeg, Manitoba, Canada}

\author{D. Pickering}

\affiliation{Department of Mathematics, Brandon University,\\
Brandon, Manitoba, R7A 6A9 Canada}

\begin{abstract}
We present the results of an analysis of superradiant energy flow due to scalar fields incident on an acoustic black hole. In addition to providing independent confirmation of the recent results in \cite{new-3}, we determine in detail the profile of energy flow everywhere outside the horizon. We confirm explicitly that in a suitable frame the energy flow is inward at the horizon and outward at infinity, as expected on physical grounds.
\end{abstract}
\maketitle

\section{Introduction}

The superradiance effect for a rotating black hole describes the situation
where fields incident on a black hole have a reflection coefficient
greater than one, and  energy is extracted from the black
hole. This theoretical prediction is extremely hard to test on remote
astrophysical objects. In order to study this and
other phenomenon of black hole physics, it is useful to
have a black hole analog which can be studied in the
laboratory. The most well known black hole analog was developed by
Unruh who showed that many aspects of black hole physics could
be reproduced in supersonic fluid flows \cite{key-1}. The acoustic
black hole (or \emph{dumb hole}) consists of fluid flowing inward
toward a fluid sink. If the rate of the fluid flow inward becomes
greater than the speed of sound in the fluid at some radial distance
from the sink, sound waves cannot escape from this \emph{event horizon}.
Since Unruh's original proposal, many different kinds of analogue black holes have been proposed \cite{new-1}. At present,
the Hawking temperatures associated with these analogue black holes are too low to be detected, but the situation is
likely to change in the near future \cite{new-2}.

In this paper we study the rotating acoustic black hole first considered in \cite{key-2} and recently studied in detail in \cite{new-3}.
The fluid is characterized by an irrotational velocity potential that has a sonic horizon and an ergoregion. The corresponding velocity field is given by:
\bea
\vec{v}=-\frac{A}{r}\,\hat{r}+\frac{B}{r}\,\hat{\phi}
\eea
 where $A$ and $B$ are positive numerical constants. We will work in dimensionless variables obtained by setting $c=A=1$ (where $c$ is the speed of sound). The solutions to the equation of motion for fluctuations in the velocity potential have the form:
\bea
\Psi(t,r,\phi)=R(r)\, e^{-i\,\omega t}e^{i\, m\,\phi}
\eea
 where $\omega$ is real and positive and $m$ must be an integer for
$\Psi(t,r,\phi)$ to be single valued. 
One defines a different radial function, $H(r)$:
 \bea
R(r)=\frac{H(r)}{\sqrt{r}}\,\exp\left[\frac{i}{2}\left\{ \omega\,\ln\left(r^2-1\right)-m \;B\,\ln\left(1-\frac{1}{r^{2}}\right)\right\} \right],
\eea
and introduces the tortoise coordinate:
\begin{equation}
r^{*}=r+\frac{1}{2}\,{\rm ln}\,\left|\frac{r-1}{r+1}\right|,
\label{tortoise_coord}
\end{equation}
which ranges from $\left[-\infty,\infty\right]$
as $r$ ranges from $\left[1,\infty\right]$ (note that $r=1$ is the event horizon). 
We obtain the differential equation \cite{new-3}:
\begin{equation}
\frac{d^{2}H(r^{*})}{dr^{*^{2}}}+\left[\left(\omega-\frac{Bm}{r^{2}}\right)^{2}-\left(1-\frac{1}{r^{2}}\right)\left\{ \frac{1}{r^{2}}\left(m^{2}-\frac{1}{4}\right)+\frac{5}{4}\,\frac{1}{r^{4}}\right\} \right]H(r^{*})=0.\label{no-A}
\end{equation}
We solve these equations numerically using the asymptotic solutions as boundary conditions. Enforcing the condition that
the only physical solution at the event horizon is the ingoing one, we obtain:
\bea
\label{soln1}
&& H(r^{*})=\mathcal{R}\, e^{i\,\omega\, r^{*}}+e^{-i\,\omega\, r^{*}}\qquad\mbox{ for }r^{*}\rightarrow\infty \\
&&H (r^{*})=\mathcal{T}\,e^{-i(\omega-m\,B)R^{*}} \qquad\mbox{ for }r^{*}\rightarrow-\infty,\nonumber
\eea
where $\mathcal{R}$ and $\mathcal{T}$ are the amplitudes of the reflected and transmitted waves and are related by the condition \cite{key-3,key-5}:
\begin{equation}
1-\left|\mathcal{R}\right|^{2}=\left(1-\frac{m\,B}{\omega}\right)\left|\mathcal{T}\right|^{2}.\label{Constant_Wronskian}
\end{equation}
For $\omega<mB$
the reflection probability $\left|\mathcal{R}\right|^{2}$ has magnitude greater than 1. Superresonance occurs
when solutions exist to (\ref{no-A}) that approach the asymptotic
forms (\ref{soln1}) for parameters that satisfy this condition.

\begin{figure}[H]
\begin{center}
\includegraphics[width=9.5cm]{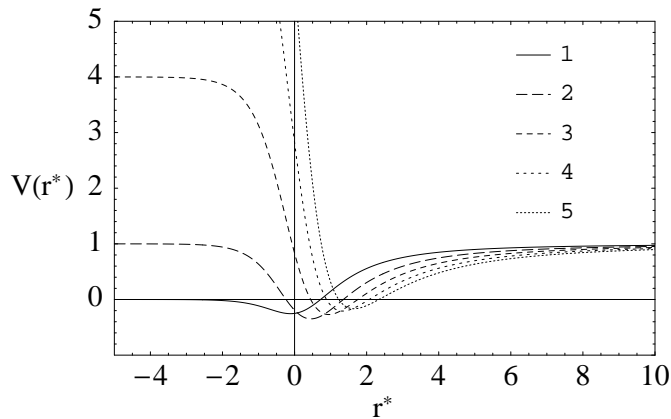}
\end{center}
\caption{Graph of V(r) vs. $r$ for values of B shown in the legend. }
\label{potential}
\end{figure}

The potential function in square brackets in (\ref{no-A}) is shown in Fig. (\ref{potential}) for different values of $B$.
\par

\begin{figure}[H]
\begin{center}
\includegraphics[width=8cm]{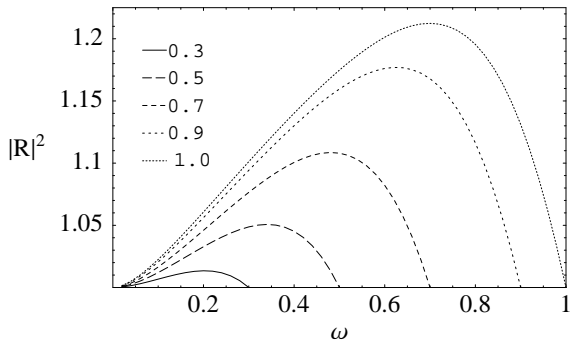}
\hfill
\includegraphics[width=8cm]{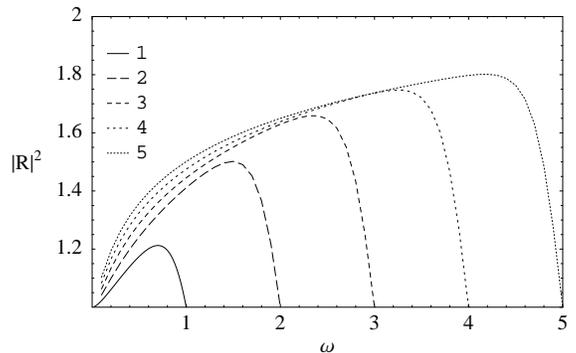}
\end{center}
\caption{Graphs of $|R|^2$ vs. $\omega$ for $m=1$ and various values of B as shown in the legends.}
\label{rsq_m1}
\end{figure}

\begin{figure}[H]
\begin{center}
\includegraphics[width=8cm]{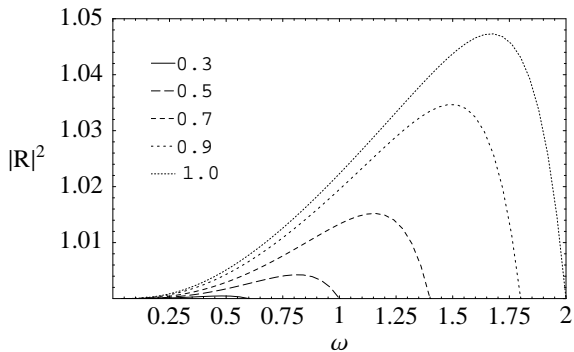}
\hfill
\includegraphics[width=8cm]{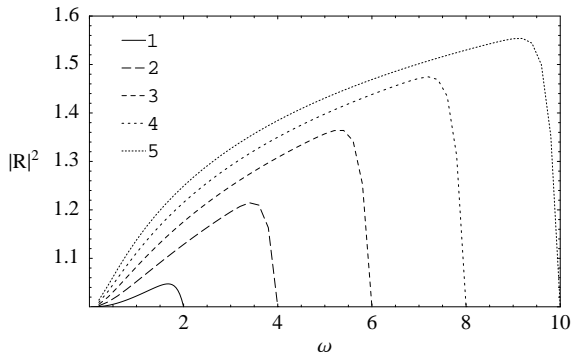}
\end{center}
\caption{Graphs of $|R|^2$ vs. $\omega$ for $m=2$ and values of B shown in the legends. }
\label{rsq_m2}
\end{figure}

Our numerical solutions to (\ref{no-A}) agree with those of \cite{new-3}. Results for the reflection coefficient are shown in Figs (\ref{rsq_m1}-\ref{rsq_m2}).
\par
\par
The superresonance effect is significant even for small
values of $B$, and can be larger than
75\% for large values of $B$. For $m=2$ the superresonant effect is still significant
but decreases with increasing $m$. Note that superresonance does not occur for negative values of $m$. 
\section{Size of the Superradiance Effect}
The size of the superresonant effect is somewhat unexpected considering that the 
superradiance of rotating astrophysical black holes is always less than $\sim$ 0.2\% \cite{key-6}. It has been pointed out in \cite{new-3} that another fundamental
difference between acoustic black holes and astrophysical black holes is that there is, in principle, no upper limit on 
the rotational velocity of the acoustic black hole. Related to this is the observation
that in the acoustic black hole there is no limit on the size of the ergoregion whereas the equatorial radius of the 
ergoregion in the Kerr black hole cannot be any bigger than twice the radius of the event horizon.  The superresonance effect, which is related to the Penrose process, 
should have some connection to the size of the ergoregion and, as we will discuss later,
the region of negative effective potential energy.  Neither of 
these observations, however, account for the difference in the magnitude of the superresonant effects. 
The superresonance effect in acoustic black holes is significant even when the radius of the ergoregion is only 
slightly greater than the event horizon.

\begin{figure}[H]
\begin{center}
\includegraphics[width=8cm]{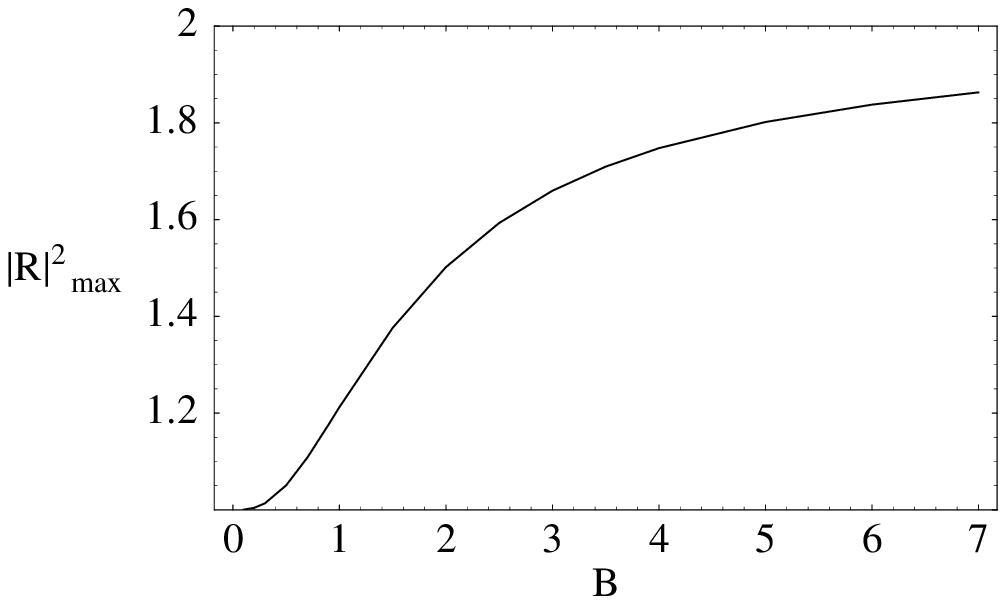}
\hfill
\includegraphics[width=7.8cm]{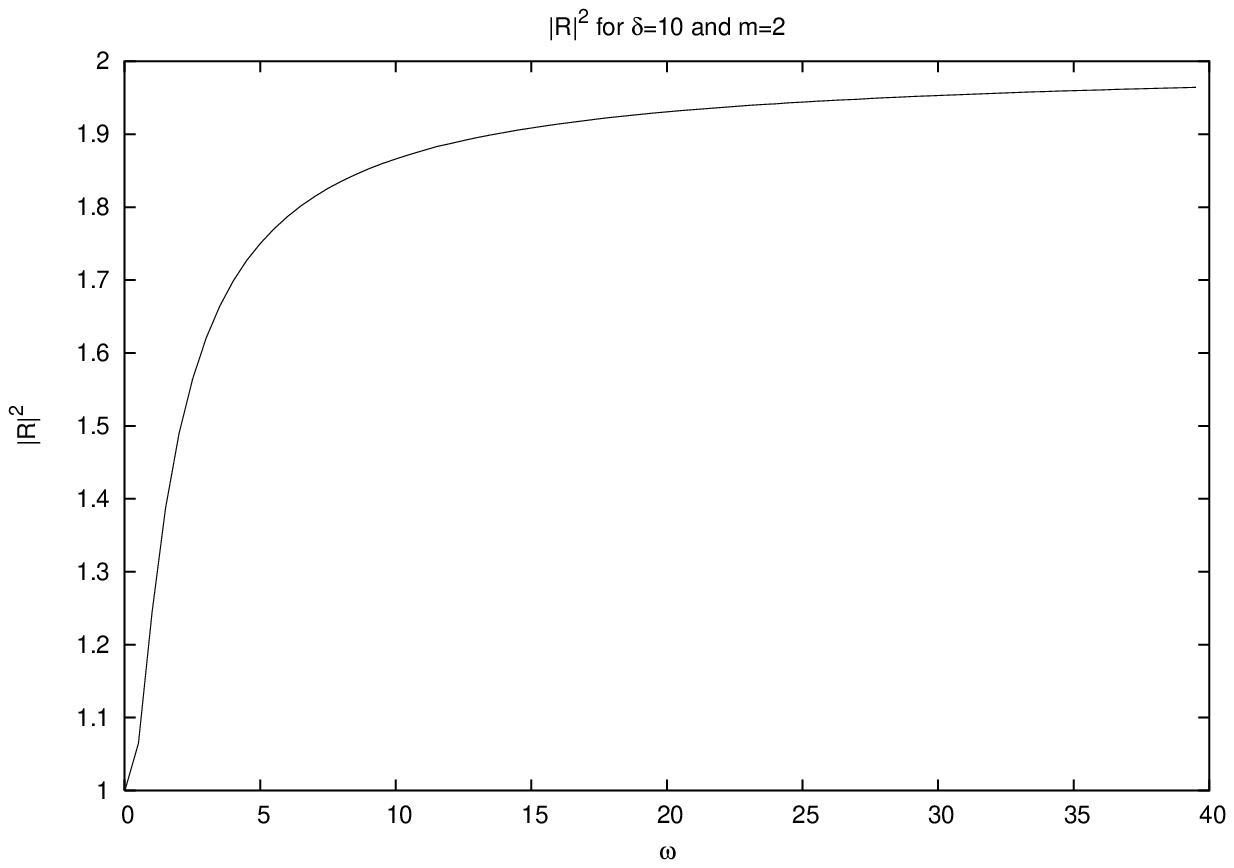}
\end{center}
\caption{The left panel is a graph of the maximum value of $|R|^2$ versus B for $m=1$ using units where $A=c=1$.  The right panel is a graph of $|R|^2$ versus $\omega$ for $m=2$ and $\delta=10$ }
\label{max_rsq}
\end{figure}

Our results indicate the existence of a maximum value of the superresonance effect as a function of B. (We note that the theory behind this analysis is not valid for arbitrarily large values of $B$. The acoustic metric is derived using hydrodynamic approximations, but if the angular component of the velocity is large enough the dispersion relation for the fluid will invalidate the assumption of a hydrodynamic model. It is estimated that the theory is valid for $B\ll 7$ \cite{new-3}). The results are shown in the left panel of
Fig. (\ref{max_rsq}).
\par
\par
\noindent  We find that $|\mathcal{R}|^2$ never rises above 2 and is a monotonically increasing function of $B$. 
\par
It is interesting to note that the existence of a limit to the size of the reflection coefficient can be seen explicitly by using a different set of variables. Defining $\delta = B/\omega$ in (\ref{no-A}) we find that the reflection
coefficient asymptotically approaches its maximum value
for fixed $m$ and $\delta$ (see right panel of Fig. \ref{max_rsq}). These variables were used in 
 the analytical analysis of Ref. \cite{key-5} (which did not consider the limit of large $\omega$).

\noindent

\par
\begin{figure}[H]
\begin{center}
\includegraphics[width=7cm]{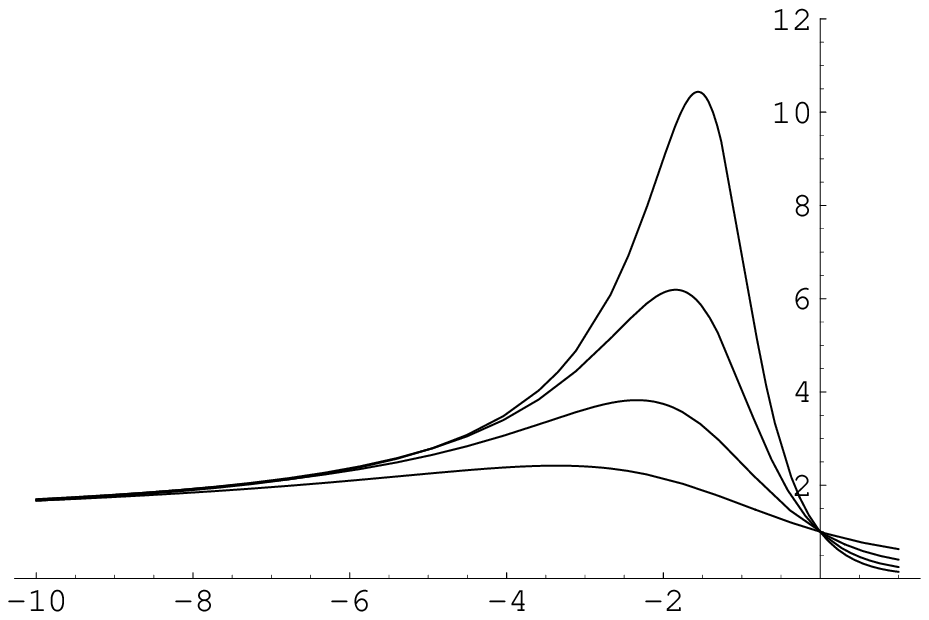}
\hfill
\includegraphics[width=7cm]{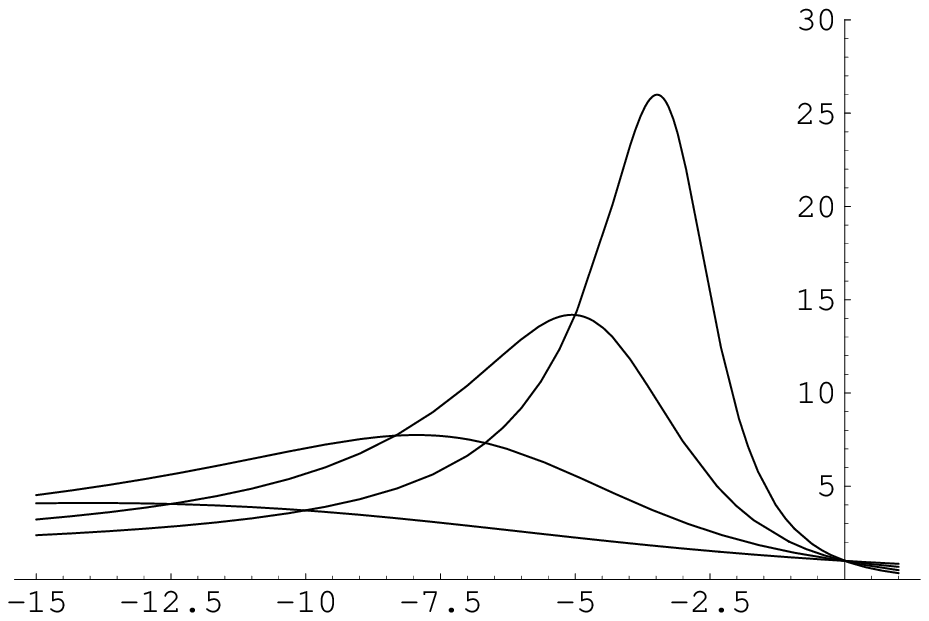}
\end{center}
\caption{Left panel is a graph of $|\mathcal{R}|^2$ versus $\alpha$ with $2\alpha\,\omega_2 = 10$. From the top the curves have $\alpha = .5 \beta$,  $\alpha = .4 \beta$, $\alpha = .3 \beta$ and $\alpha = .2 \beta$.  Right panel is same graph with $2\alpha\,\omega_2 = 5$.}
\label{RK}
\end{figure}

Ignoring for the moment the restrictions imposed by hydrodynamics, there is no apparent reason that $|\mathcal{R}|^2_{max}$ cannot be arbitrarily large, for arbitrarily large values of $B$. It is interesting to verify that one can reproduce this asymptotic behaviour in simple model that can be solved analytically.  
 We consider the general equation
\begin{equation}
\frac{d^2H}{dx^2} +V(x)H=0
\end{equation}
and use a potential of the form:
\begin{eqnarray}
V(x) = \left\{
\begin{array}{cc}
\omega_1^2 & x < -a \\[2mm]
\omega_2^2 & -a < x < a \\[2mm]
\omega_3^2 & x > a 
\end{array}
\right. ~~~~~~\omega_3 = \alpha \omega_1; ~ \omega_2 = \beta \omega_1
\end{eqnarray}
We use a solution of the form 
\begin{eqnarray}
H(x) &=& A\exp(i\omega_1 x) + B\exp(-i\omega_1 x) \qquad x<-a\nonumber\\
&=& C\exp(i\omega_2 x) + D\exp(-i\omega_2 x) \qquad -a<x<a\nonumber\\
 &=& E\exp(i\omega_3 x) \qquad x>a\nonumber\\
\end{eqnarray}
and demand continuity of $H(x)$ and it's derivative at $x=\pm a$ to obtain $|R|^2 + \alpha|T|^2 = 1\nonumber
$ and
\begin{eqnarray}
&&|R|^2 = \frac{
(1-\alpha)^2\cos^2(2\omega_2 a) +
(\beta-\frac{\alpha}{\beta})^2\sin^2(2\omega_2 a) }
{(1+\alpha)^2\cos^2(2\omega_2 a) +
(\beta+\frac{\alpha}{\beta})^2\sin^2(2\omega_2 a) }\nonumber\\
&&|T|^2 = \frac{4 }
{(1+\alpha)^2\cos^2(2\omega_2 a) +
(\beta+\frac{\alpha}{\beta})^2\sin^2(2\omega_2 a) }\nonumber\\
&&|R|^2 + \alpha|T|^2 = 1\nonumber
\end{eqnarray}
For $\beta\ne 1$  superresonnance occurs for $\alpha < 0$. The analytic results for this simple model clearly indicate the existence of a maximum value for $|R|^2$ (see Fig. \ref{RK}), in analogy with the numerical results for the acoustic black hole. Further analysis is required to determine whether or not this is a generic property of superresonance.
\par

\section{Energy Flow}
Our numerical results, obtained concurrently with those of \cite{new-3}, confirm the existence of superresonnant solutions for acoustic black holes. If these mathematical solutions correspond to a 
physical effect, it should be possible to verify that there is a net outward flow 
of energy from the black hole whenever the condition $|{\cal R}|^2 > 1$ is satisfied.  Moreover, for both superresonnant and non-superresonnant solutions, we should find that the energy flow asymptotically close to the event horizon is inward, since energy should not be able to escape from within the event horizon. These points are well understood in astrophysical black holes \cite{key-8} but have not been studied in detail in the present context.
The purpose of this section is to verify that  the numerical solutions to the differential equations found in this paper and in  \cite{new-3} do indeed give rise (in a suitable frame) to energy flow that is outward in the region asymptotically far from the event horizon, and inward in the region asymptotically close to the event horizon, as expected on physical grounds.

Consider the asymptotic form of the solution very close to the horizon. From (\ref{soln1}) we have:
\bea
H(r^{*})=\mathcal{T}\,\exp\left\{ i\,\sigma \, r^{*}\right\} \qquad\mbox{ where }\sigma=-(\omega-m\,B),
\eea
In the region of parameter space where superresonance occurs ($\frac{m\,B}{\omega}>1$) we have $\sigma >0$ and thus, according to an observer at infinity, the solution at the event horizon is 
an outgoing wave with phase velocity given by: 
\bea
v_{phase}=\frac{\sigma}{\omega}=-(1-\frac{m\,B}{\omega}).
\label{phase_velocity}
\eea
Naively, this result might suggest a connection between a positive outgoing phase velocity 
asymptotically close to the event horizon, and the positive energy extracted from a superresonnant acoustic black hole (see Appendix B of 
\cite{key-9}). 
A similar result was found in \cite{key-6} where it was claimed that, for a light pulse with superradiant 
frequency incident on a rotating astrophysical black hole, the energy flowing out of a surface very near the event
horizon is positive and only slightly less than the energy flowing out of a surface very far from the black hole.  However, as mentioned above, outward energy flow arbitrarily close to the event horizon essentially corresponds to energy 
coming out of the event horizon. Thus, this naive argument, and the results of \cite{key-6}, seems to suggest unphysical solutions. 

However, as is well known, the direction of the energy flow 
is governed by the group velocity, and not the phase velocity.  In this case the group velocity is given by \footnote{This
analysis closely parallels that given in \cite{key-8} on pg. 427.}:
\bea
v_{group}=\frac{d\sigma}{d\omega}=-1
\eea
which means that, at the event horizon, energy is flowing inward at the speed of sound, in apparent contradiction to statements made in \cite{key-6}  (see endnote \footnote{This apparent outward flow of energy from the event horizon is an artifact of
the choice of frame and does not arise in the locally nonrotating frame defined in \cite{key-9}.}).

In order to analyse the energy flow of our solutions,  we note that the energy flux across
an arc $dS$ of a circle of constant $r$ with unit normal vector $r^\nu$  is given by:
\bea
dE=T^{\alpha\beta} t_\alpha r_\beta\,\, dS=T^{0r}\, dS.
\eea
where $T^{\alpha\beta}$ is the stress-energy tensor of the massless scalar field and is 
given by:
\begin{equation}
T^{\alpha\beta}=g^{\mu\alpha}g^{\nu\beta}T_{\mu\nu}\;;~~T_{\mu\nu}=\frac{1}{2}\left(\nabla_{\mu}\Psi^{*}\,\nabla_{\nu}\Psi+\nabla_{\mu}\Psi\,\nabla_{\nu}\Psi^{*}\right)-\frac{1}{2}\, g_{\mu\nu}\nabla_{\lambda}\Psi\nabla^{\lambda}\Psi^{*}.
\end{equation}
In terms of the numerical solutions for $H(r^*)$ the $T^{0r}$ component of the stress energy tensor is:
\begin{eqnarray}
T^{0r}&=&\frac{\left(1-4m^2-\frac{1}{r^2}-\frac{4(B m-r^2\omega)}{r^2-1}\right)H(r^*)H^*(r^*)}{8r^4\rho_0^4}
+\frac{H^\prime(r^*)\left(H^{\prime}(r^*)\right)^*}{2 (r^2-1) \rho_0^4} \nonumber \\
&-& \frac{i(B m-r^2\omega)\left\{  H(r^*)\left(H^{\prime}(r^*)\right)^*-H^\prime(r^*)H^{*}(r^*) \right\}}{2r(r^2-1)\rho_0^4} 
-\frac{ H(r^*)\left(H^{\prime}(r^*)\right)^*+H^\prime(r^*)H^{*}(r^*) }{4r^3\rho_0^4} 
\end{eqnarray}

\begin{figure}[H]
\begin{center}
\includegraphics[width=7cm]{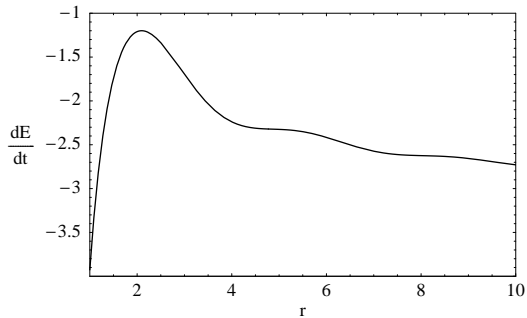}
\end{center}
\caption{ Graph of the rate of outward energy flow across a circle of radius $r$  surrounding the acoustic black hole for $B=0.5$, $m=1$ and $\omega=1$. The energy flow is strictly negative indicating that net energy flow is inward everywhere.}
\label{energyflow_nonsuperresonant_short}
\end{figure}

We use the numerical solutions of the differential equation to study the energy flow in both superresonant 
and non-superresonant cases.  In non-superresonant cases we find that, as one would expect, the energy flow 
is inward for all values of $r$. An example is shown in Fig. 
(\ref{energyflow_nonsuperresonant_short}).

\begin{figure}[H]
\begin{center}
\includegraphics[width=7cm]{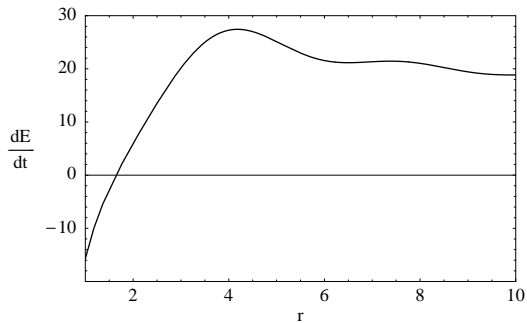}
\end{center}
\caption{Graph of the rate of outward energy flow across a circle of radius $r$  surrounding the acoustic 
black hole for $B=1$, $m=1$ and $\omega=1$. Far from the event horizon the rate of outward energy
flow is positive which indicates superresonance.  The energy flow near the horizon ($r=1$) is negative 
indicating that the energy flow is inward. }
\label{energyflow_superresonant_short}
\end{figure}

In superresonant cases we find that the energy flow is inward in the region near the horizon and outward  far from the horizon (see Fig. \ref{energyflow_superresonant_short}). 

As a check of the numerical 
results, we have confirmed that in the limit $r\rightarrow 1$, the numerical result for the energy flow approaches the result that one obtains analytically using the 
asymptotic solution (\ref{soln1}).  In both cases the energy flow asymptotically 
approaches a constant value as $r\rightarrow \infty$.



\begin{figure}[H]
\begin{center}
\includegraphics[width=8cm]{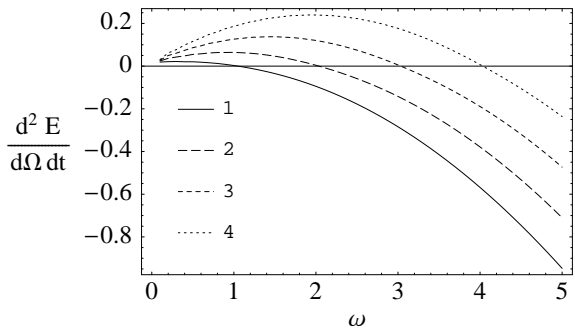}
\end{center}
\caption{ Graph of outward energy flow through a circle at $r=20$ as a function of $\omega$ 
for $m=1$ and different values of $B$. As $\omega$ increases, net outward energy flow ceases in each case very close
to the cutoff frequency ${\omega} = m\,B$, for the superresonance effect.}
\label{energyflow_omega}
\end{figure}

We have also studied how the energy flow, at a fixed value of $r$ far from the black hole,
depends on the frequency $\omega$ of the incident wave. The results are shown in 
Fig. (\ref{energyflow_omega}). 
\noindent Comparing this graph with the rightmost graph in Fig. (\ref{rsq_m1}) we can see that for each value of $B$, the range of $\omega$ for which the energy flow is positive is almost exactly equal to the range of $\omega$ for which $|{\cal R}|^2>1$ (which is given by $\omega < m \,B$).
We note that the value of $\omega$ for which the energy flow changes sign approaches $\omega = m \,B$, in the limit that we look at the energy flow asymptotically far from the acoustic
black hole.  At finite radius, the frequency at which the direction of the energy flow changes  is 
always bigger than $\omega = m \,B$.

\begin{figure}[H]
\begin{center}
\includegraphics[width=9.5cm]{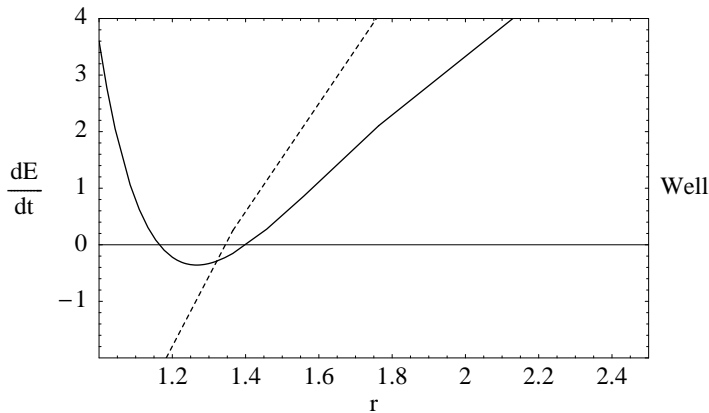}
\end{center}
\caption{Graph of the well function for  $B=1.6$, $m=1$ and $\omega=1$ (solid line), and the rate of energy flow as a function of $r$ for same parameters (dotted line).
The net energy flow becomes outward at a radius greater than the smallest value of $r$ for which  the potential crosses zero.}
\label{energyflow_well}
\end{figure}

The radius for which the energy flow becomes positive can be related to a parameter of the well function.
From  Fig. (\ref{potential}) it can be seen that the well function contains a small negative region. We call the smallest value of $r$ for which the potential crosses zero $r_1$, and the larger value of $r$ where the potential becomes positive again is called $r_2$. In all cases, the  energy flow becomes positive at $r > r_1$. This result is shown in Fig. (\ref{energyflow_well}).  
This result is
reminiscent of the Penrose process, where energy can be emitted by an astrophysical
black hole by creation of back to back photons in a region where the effective 
potential for particles travelling opposite to the rotation of the black hole is 
negative.  In the present case, the location of the transition between outward and inward energy
flow satisfies $r>r_1$, but not $r_2>r>r_1$. 

These results confirm  that there is a net outward flow of energy from a rotating acoustic 
black hole in the superresonant regime of parameter space. The energy originates a finite 
distance outside the event horizon,  and the energy flow close to the event horizon is inward.  
We conclude that the numerical
superresonant solutions obtained in our analysis and that of \cite{new-3} are physical solutions.

\section{Conclusion}

 To summarize, we have obtained, concurrently and in agreement with the authors of  
\cite{new-3}, a
maximum superresonance effect that corresponds to an increase in the reflected amplitude with 
respect to the incident amplitude that is on the order of 100\% (apart from any limits imposed by hydrodynamics).
An analytic solution of a one dimensional toy model with similar properties is also shown
to exhibit such a maximum.  
By analysing the  energy flow, we have explicitly shown that the properties of the energy flow arising in the numerical superresonant 
solutions are consistent with physical expectations.

\vspace{0.2in}

\leftline{\bf Acknowledgments}
This research was supported in part by the Natural Sciences and
Engineering Research Council of Canada.

\end{document}